\documentclass[aps,prl,floatfix,superscriptaddress,twocolumn,footinbib]{revtex4-1}
\usepackage{graphicx}% Include figure files
\usepackage{dcolumn}% Align table columns on decimal point
\usepackage{bm}% bold math
\usepackage{makecell}
\usepackage{bbold} %% 用于显示\mathbb{1}
\usepackage{color}
\usepackage{amsmath}
\usepackage{latexsym}
\usepackage{amssymb}
\usepackage{graphics,epstopdf}
\usepackage{hyperref}
\usepackage{xcolor}
\hypersetup{
    colorlinks = true,
    linkcolor =blue,
	citecolor=blue,
	urlcolor=blue
}

\newcommand{\ed}{\end{document}}
\newcommand{\beq}{\begin{equation}}
\newcommand{\eeq}{\end{equation}}

\usepackage{mathtools}
\usepackage{comment}

\begin{document}

%\title{Chirality-Induced Majorana Polarization in Circular Helix Molecules}
\title{Chirality-Induced Majorana Polarization}

\author{Song Chen}
 \affiliation{School of Physics and Wuhan National High Magnetic Field Center,
Huazhong University of Science and Technology, Wuhan 430074, People's Republic of China.}

\author{Hua-Hua Fu}
\altaffiliation{Corresponding author.\\ hhfu@hust.edu.cn}%Lines break automatically or can be forced with \\
\affiliation{School of Physics and Wuhan National High Magnetic Field Center,
Huazhong University of Science and Technology, Wuhan 430074, People's Republic of China.}
\affiliation{Institute for Quantum Science and Engineering, Huazhong University of Science and Technology, Wuhan, Hubei 430074, China.}

\date{\today}

\begin{abstract}
To realize Majorana fermions having novel physical features has been developed as a key while difficult task in topological superconductor. Here we have proposed another platform to generate Majorana zero modes (MZMs), which is constructed by a single opened circular helix molecules (CHM) coupled with a s-wave superconductor (with magnetic field) or by an interlinked-CHMs chain coupled with a phase-bias s-wave superconducting heterostructure (without any magnetic field). The MZMs achieved here are tightly associated with the structural chirality in CHMs. Importantly, the left and right handedness may result in completely opposite Majorana polarization (MP), and the local MP is associated to the chiraliy-induced spin polarization. These properties provides us multiple effective ways to detect and regulate the MZMs by using the chirality-induced spin selectivity (CISS) effect and the related spin-polarized currents in chiral materials. 
                
\end{abstract}

\maketitle

{\color{blue}\emph{Introduction.}}
Majorana fermions, novel quasiparticles emerging in topological superconductors, have been attracting increasing research interest due to their promising applications in topological quantum computing because of their unique non-Abelian exchange statistics \cite{Nayak2008,Oreg2020,Beenakker2013,Zhang2024}. A pioneering proposal given by Kitaev demonstrated that the isolated Majorana zero modes (MZMs) could exist at the ends of one-dimensional spinless $p$-wave superconductors \cite{Kitaev2001}, and then both experimental and theoretical efforts have been continuously advancing to realize them in real materials. Owning to the advent of quantum matter in recent decays \cite{Liu2020,Wang2022PRB,Yang2024}, another groundbreaking proposal that a topological insulator coupled with a standard s-wave superconductor may serve as a platform to generate MZMs has triggered a research frenzy on this topic \cite{Fu2008}. Subsequently, some other material platform with unique physical mechanisms, such as  semiconductor nanowires \cite{Lutchyn2010,Oreg2010,Deng2012,Mourik2012,Finck2013}, and topological insulator in-plane Zeeman field \cite{Pan2019,Pan2022,Wu2020,Wang2022}  in proximity to superconductors, phase-biased Josephson junctions \cite{vanHeck2014,Lesser2021PRB, Potter2013,Fornieri2019,Ren2019,  Lesser2021}, nanowires of magnetic atoms \cite{NadjPerge2014,Ruby2017,Feldman2016}, and carbon nanotubes \cite{Marganska2018,Lesser2020} placed on superconducting (SC) substrates, have been designed for achieving these magical topological quasiparticles. Obviously, to establish an innovative material platforms is still the most critical issue to realize MZMs with novel physical features.

We well know that chirality-induced spin sepectivity (CISS) is a fascinating effect where electrons get spin polarized after propagating through organic chiral molecules \cite{Gohler2011, Xie2011,1.1,1.2,1.3,1.3.1, 1.5,1.6,1.7,1.8,1.9,1.12,1.13,1.14,Zollner2020, Hoff2021,Naskar2023} and inorganic chiral crystals \cite{Hu2024, Inui2020} such as oxides \cite{Widmer2006} and perovskites \cite{Ma2021} without applying external magnetic field, making the nonmagnetic chiral molecules have already been a star material to realize various inspiring spin-associated phenomena, such as the paired opposite polarized spin polarity in superconductors \cite{Nakajima2023} and the CISS-driven anomalous Hall effect \cite{Sinova2004,Valenzuela2006}. Furthermore, in chiral molecules-superconductors hybrid systems, zero-bias conductance peak (ZBCP) are observed \cite{Kasumov2001,Alpern2016,Shapira2018,Alpern2019}. These peaks reduce but do not split in a magnetic field, indicating equal-spin triplet superconductivity with either even-frequency p-wave or odd-frequency s-wave symmetries \cite{Shapira2018,Alpern2019}.  However, previous studies on these hybrid systems have primarily focused on linear open chiral molecules, potentially limiting the exploration of novel physical phenomena \cite{Tang2019, Chen2023}. For instance, in our earlier works, the spin destructive quantum interference  and PCISS effect was observed in circular helical molecules (CHMs) due to the unique spin Berry phase of the system \cite{ChenFu2023,ChenFu2024}. This new mechanism inevitably provide us another opportunity to realize some novel quantum states including the magic MZMs.

% In our previous work, we have uncovered a new class of CISS effect, i.e., the persistent CISS (PCISS) effect in circular helix moelcues (CHMs), which extends the conventional CISS effect occurring in the non-equilibrium transport process to the equilibrium transport process for the first time. The PCISS effect combined with other unique advantages in CHMs, such as the chirality-induced spin-orbit coupling (SOC), the quantum coherence, the ring-induced Berry phase in the absence of magnetic field, without applying external metallic electrodes and other, inevitably provide us another opportunity to realize some novel quantum states including the magic MZMs.         

In this work, we have established theoretically a new while effective topological superconductor (SC) to realize MZMs. The topological SC platform is constructed by a open CHM coupled with a $s$-wave SC (with an external magnetic field) or a series of interlinked-CHMs in proximity to a phase-bias $s$-wave SC heterojunction substrate (without any external magnetic field). In this model design, the Aharonov-Casher (AC) phase induced by the SOC inherent in CHMs and the phase winding in the SC substrates provide the prerequisite for the occurrence of topologically nontrivial phases in some phase spaces. Our theoretical results demonstrate the emergence of a pair of MZMs at both ends of the open CHM and the interlinked-CHM chain. More importantly, this kind of MZMs are tightly associated with the structural chirality in CHMs. Particularly, we find that the left and right handedness may result in completely opposite Majorana polarization (MP). Considering this kind of chirality-associated MZMs have not reported anywhere previously, we refer to this phenomenon as the chirality-induced MP (CIMP). Moreover, the local CIMP is also associated  to the chiraliy-induced spin polarization. This inspiring property provides us an effective way to detect the MZMs by a spin-polarized density of state measurement and to regulate the MZMs by the CISS effect in chiral materials.

\begin{figure}
\includegraphics[width=\columnwidth]{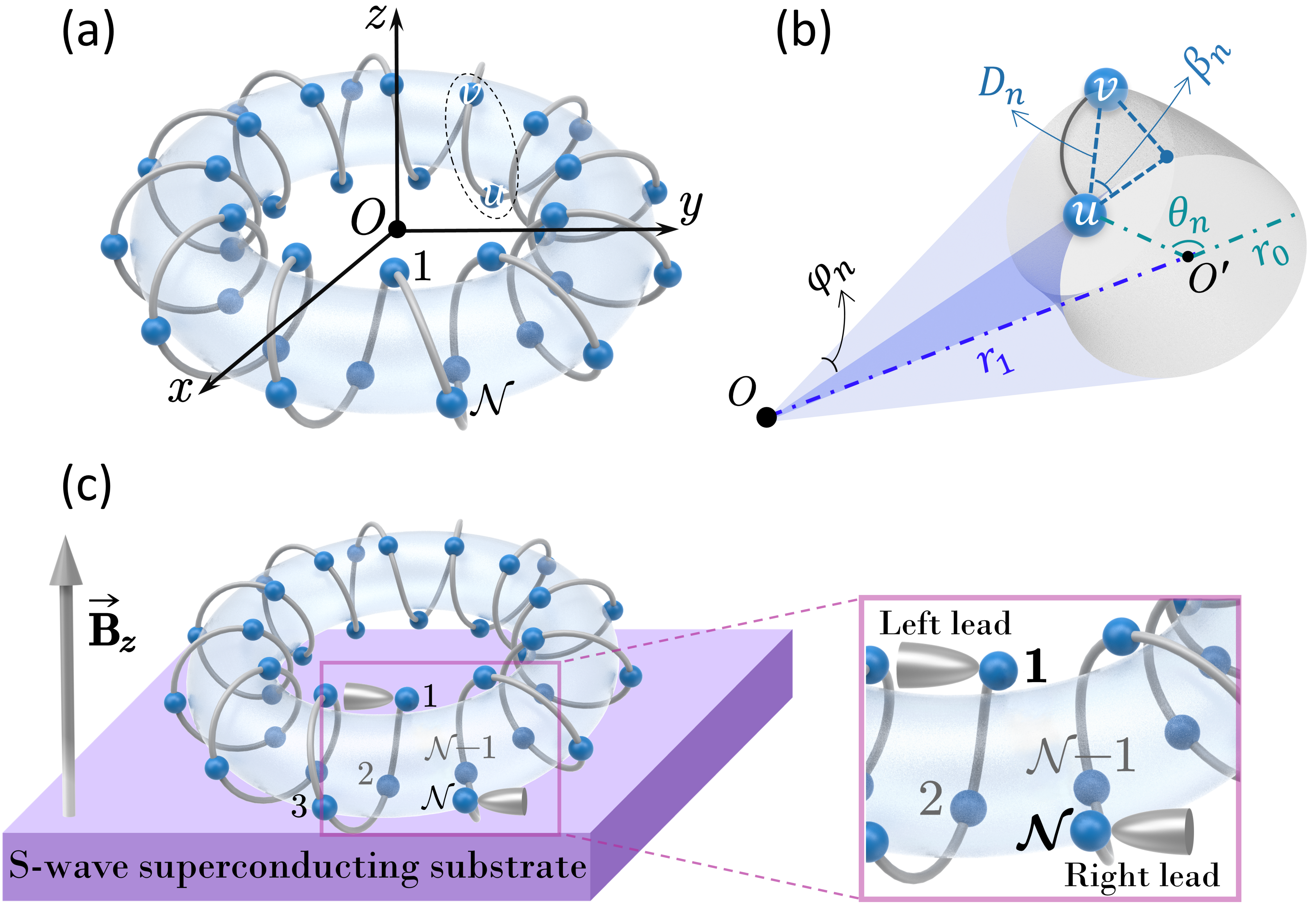}
\caption{\textcolor{black}{(a) Side schematic view of a single CHM. (b) A magnified schematic of the portion enclosed by the black dashed in (a). Here $O$ ($O^{\prime}$) and $r_1$ ($r_0$) denote the center and radius of the toridal chiral molecule and of its crossing-section plane. $\phi_n$ and $\theta_n$ indicate the toroidal and poloidal angle. $\beta_n$ and $D_n$ represent the space angle and the distance between two neighbouring atoms in the above plane. (c) A broken CHM and its 1st and $\mathcal{N}$th lattice cites are adsorbed onto an $s$-wave SC substrate (purple region) with a Zeeman field along the $z$-direction. An enlarged diagram shows the 1st and $\mathcal{N}$th lattice cites coupled with two nonmagnetic electrodes.}}
\label{fig1}
\end{figure}

{\color{blue}\emph{Device model.}} To achieve isolated MZMs in this quasi-1D structure based on CHM, we break the connection between the first and the $\mathcal{N}$th lattice sites in CHM (see Fig. \ref{fig1}(a)) and couple them to an $s$-wave SC substrate, where time-reversal symmetry is broken by applying a magnetic field in the $z$-direction, as illustrated in Fig. \ref{fig1}(c). To gain the transport properties along the CHM, we adopt two nonmagntic metal leads to couple with the first and the $\mathcal{N}$th lattice sites by using thiol group-mediated chemical adsorption \cite{Xie2011}. In the absence of SC substrate, the broken CHM can be described by the following Hamiltonian
\begin{equation}
\begin{aligned}
\mathcal{H}_{c}  = &\sum_{n=1}^{\mathcal{N}} \varepsilon_n c_{n}^\dagger c_{n} + \sum_{n=1}^{\mathcal{N}} B_z c_{n}^\dagger \sigma_z c_{n} + \sum_{n=1}^{\mathcal{N}-1} t_{n} c_{n}^\dagger c_{n+1}  \\
&+ \sum_{n=1}^{\mathcal{N}-1}  2is\cos (\theta_{n}^{-}) \cos(\varphi_{n}^{-}) \sigma_{n} c_{n}^\dagger c_{n+1} + \text{H.c.},
\label{eq1}
\end{aligned}
\end{equation}
where $c_{n}^{\dagger}=(c_{n \uparrow}^{\dagger}, c_{nl \downarrow}^{\dagger})$ and $\varepsilon_{n}$ are the creation operator and the on-site energy for electrons at the $n$th lattice wiht $\mathcal{N}$ the total number of lattices in CHM and $B_z$ the Zeeman field. Moreover, $t$ and $s$ denote the electronic hopping integral and the strength of SOC respectively, since the nearest-neighboring (NN) hoppings are considered. $\sigma_n=(\sin \varphi_{n}^{+}\cos \beta _{n}$-$\sin \theta _{n}^{+}\cos \varphi _{n}^{+}\sin \beta _{n}) \sigma_x-(\sin \theta _{n}^{+}\sin \varphi _{n}^{+}\sin \beta _{n}$+$\cos \varphi _{n}^{+}\cos \beta _{n}) \sigma_y+\cos \theta _{n}^{+}\sin \beta _{n} \sigma_z$, with $\theta _{n}^{\pm}$=$(\theta_{n+1} \pm \theta _n )/2$, $\theta _n$=$(n-1)\Delta \theta$; $\Delta \theta$=$2\pi /\mathcal{M}$, and $\varphi _{n}^{\pm}$=$(\varphi _{n+1} \pm \varphi _n ) /2$, $\varphi_n$=$(n-1)\Delta\varphi$, $\Delta \varphi$=$2\pi/\mathcal{N}$ with $\mathcal{M}$ the total number of atoms in each unit cell. The space angle $\beta_{n}$ is defined by $\beta_{n}$=$\mathrm{arc}\cos \left[ X_{n}/D_{n} \right]$, where $X_{n}$=$2 r_0 \sin \left[\Delta \theta/2\right]$ and $D_{n}$ represents the distance between two neighbouring lattices and can be described by the related distance formula \cite{ChenFu2024}.

When the CHM is placed on an $s$-wave SC substrate, an on-site Cooper pairing potential $\Delta$ should be induced in the adjacent lattices. Considering furhter the superconducting pairing, the system is characterized by 
\begin{equation}
\begin{aligned}
\mathcal{H}^{\text{C}} = \mathcal{H}_{c}+  \sum_{n,l} \mu c_{nl}^\dagger c_{nl} + (\Delta_{n} c_{nl\uparrow }^\dagger c_{nl\downarrow }^\dagger + \text{H.c.}),
\label{eq2}
\end{aligned}
\end{equation}
here $\mu$ denotes the chemical potential. To determine the spectrum of a superconducting CHM, we reform $\mathcal{H}^{\text{C}}$ in a particle-hole symmetric form by introducing a Nambu spinor, this is, $\Psi = \bigoplus_{n=1}^\mathcal{N} \Psi_n$ and $\quad \Psi_n^\dagger = \left( c_{n\uparrow}^\dagger, c_{n\downarrow}^\dagger, c_{n\uparrow}, c_{n\downarrow} \right)$, where $\bigoplus$ denotes the direct sum over the $\mathcal{N}$ lattice positions, which effectively doubles the system's degrees of freedom. Then the Bogoliubov-de Gennes (BdG) Hamiltonian is described by $\mathcal{H}_{\text{BdG}} = \frac{1}{2} \Psi^\dagger \mathbf{H}_{\mathrm{BdG}} \Psi$ as below
\begin{equation}
\mathbf{H}_{\mathrm{BdG}} =
\begin{pmatrix}
\mathcal{H}_{c} + \mu & -i \Delta \sigma_y \\
i \Delta^* \sigma_y & -\mathcal{H}_{c}^* - \mu
\label{eq3}
\end{pmatrix}.
\end{equation}
Additionally, the Hamiltonian for the two metal leads is given by $\mathcal{H}^{\text{L}}$ $=\sum_{k, \beta}(\varepsilon_0 a_{\beta k}^{\dagger} a_{\beta k}+t_0 a_{\beta k+1}^{\dagger} a_{\beta k}+\text { H.c.})$, and the coupling between both leads and the CHM can be written as $\mathcal{H}^{\text{LC}}$ $=\sum_{\beta}(\gamma_{\beta}a_{\beta1}^{\dagger}c_{n_{\beta}}+\mathrm{H}.\mathrm{c}.)$, where $\beta$=$L$, $R$, $n_L$=1, $n_R$=$\mathcal{N}$, $\epsilon_0$ the on-site energy, $t_0$ the hopping strength and $a_{\beta k}^{\dagger}$ the creation operations of the $k$th position in leads. Note that $\mathcal{H}^{\text{LC}}$ plays a vital role in facilitating the charges' transferring in device. 

{\color{blue}\emph{Topological Phase Diagram.}} To perform our studies, a representative CHM is adopted and its two key structural parameters are adopted as $r_0 = 7$$\textup{~\AA}$ and $r_1 =\mathcal{N}h/2\pi$ with $\mathcal{N} = 200$ and $h=3.4$$\textup{~\AA}$. Moreover, $\varepsilon_n$ and $t$ are set as zero and $0.1 \, \text{eV}$, thus the SOC is estimated to be $s=10\, \text{meV}$. In addition, the chemical potential $\mu_0$, the SC order parameter $\Delta$ and the Zeeman energy $B_z$ are set as $150 \, \text{meV}$, $5 \, \text{meV}$ and $8 \, \text{meV}$, respectively. For real leads, $\Gamma = 5 \, \text{meV}$ are adopted, unless otherwise stated. 

\begin{figure}[t]
\includegraphics[width=\columnwidth]
{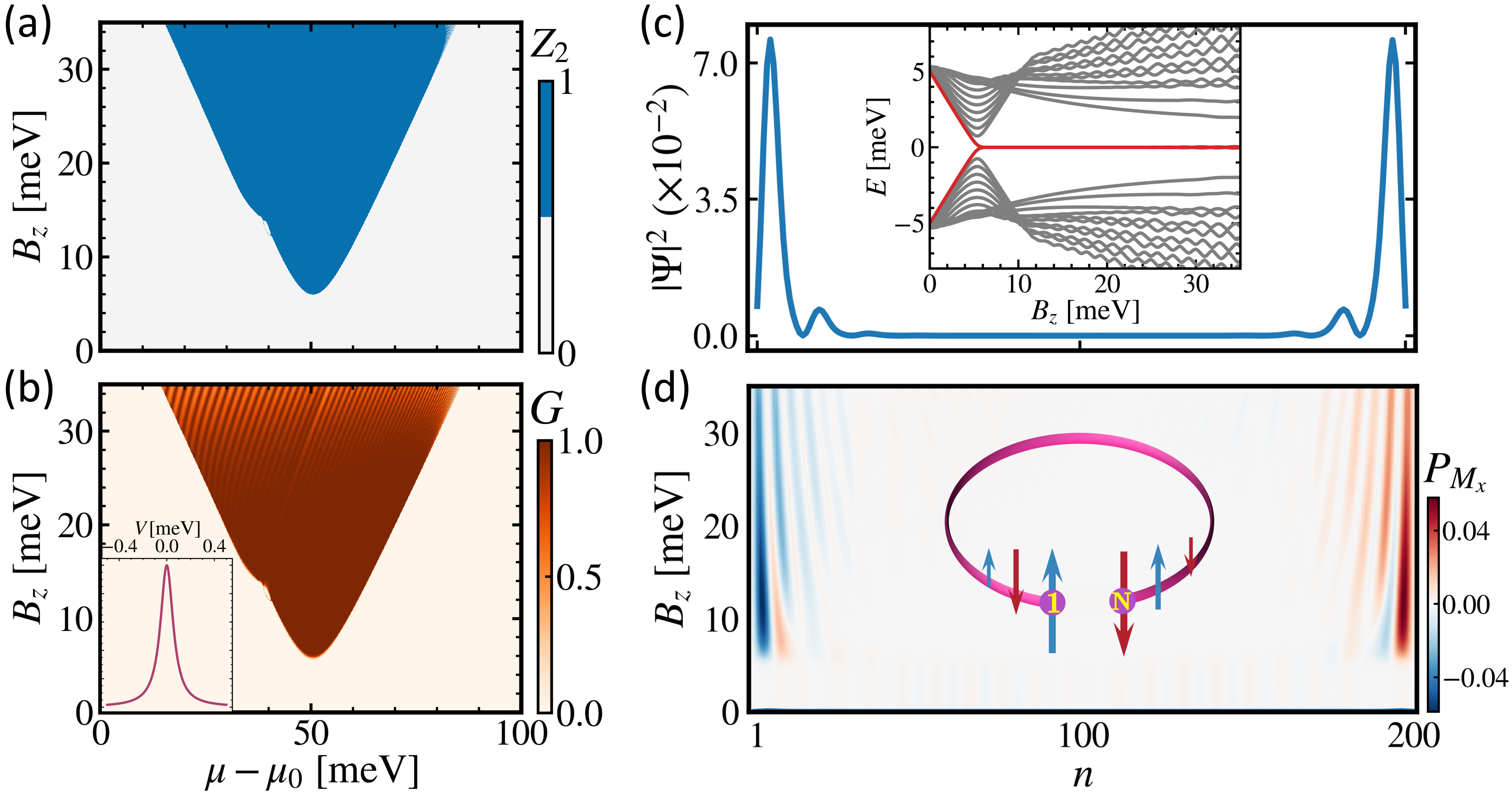}
\caption{\textcolor{black}{(a) Topological Phase diagrams by the $Z_2$ umber versus $B_z$ and $\mu$. (b) ZBCP versus $B_z$ and $\mu$, where the differential conductance $G$ versus the bias voltage ($V$) is also shown. (c) Real space distribution of the MZM wave function $|\Phi|^2$ with $B_z = 8 \, \text{meV}$ and $\mu = 50 \, \text{meV}$. An inset shows the lowest BDG energies versus $B_z$. (d) MP versus $B_z$ and the lattice position $n$ with $\mu = 50 \, \text{meV}$. Other parameters are adopted as $\mathcal{N} =200$, $t = 0.1 \, \text{eV}$,} $s=10\, \text{meV}$, $\mu_0 = 150 \, \text{meV}$, $\Delta = 5 \, \text{meV}$ and $\Gamma = 5 \, \text{meV}$.}
\label{fig2}
\end{figure}

To determine the topological properties of the CHM coupled by an $s$-wave SC, we present the topological phase diagram by examining the topological invariant $Z_2$ number as a function of $B_z$ and $\mu$ as shown in Fig. \ref{fig2}(a), where the blue region represents the topological nontrivial phases hosting MZMs. It is important to stress that the topological phase persists over a substantial range of parameters, indicating that the system does not require fine-tuning to support MZMs. This robustness to variations is crucial for experimental fabrication and measurement, supporting the CHMs as an excellent platform for realizing MZMs. Particularly, the zero-bias conductance peak versus the parameters ($B_z$, $\mu$) (see Fig. \ref{fig2}(b)) is well consistent with the above nontrivival phase diagram, providing a potential experimental evidence for the existence of MZMs.    

\begin{figure*}
\includegraphics[width=1.9\columnwidth]{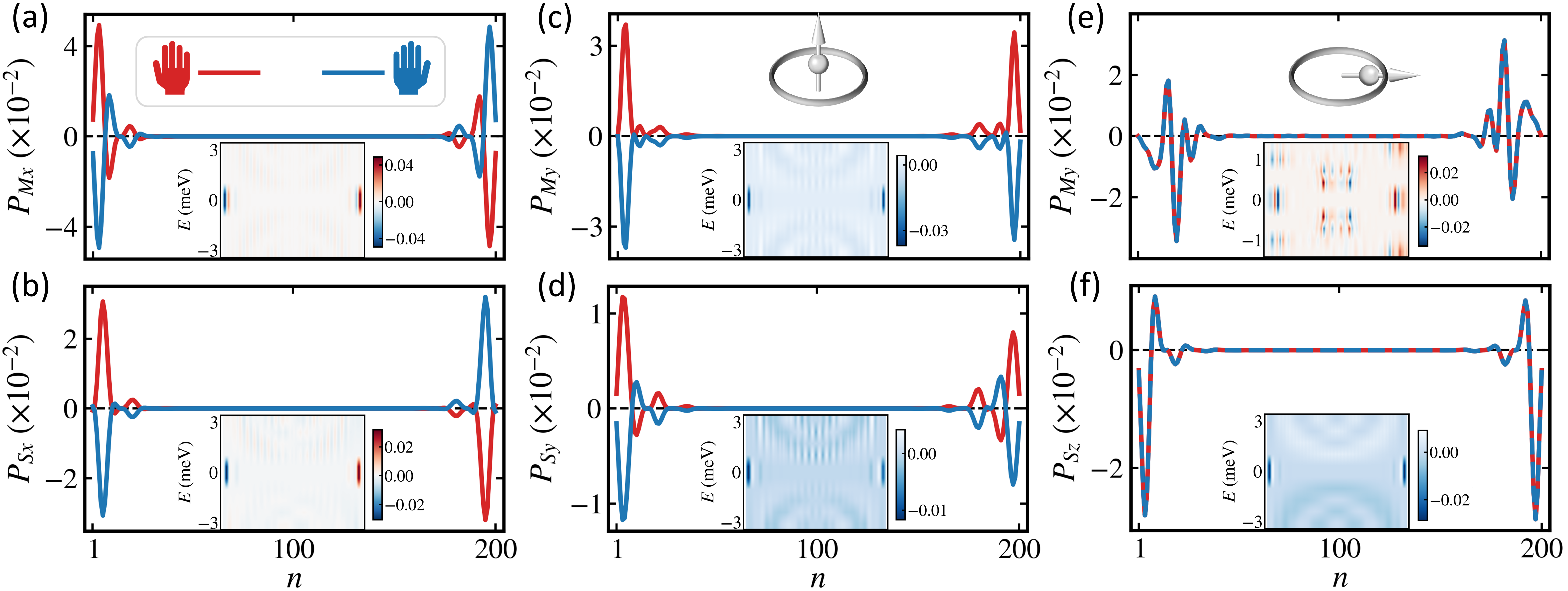}
\caption{\textcolor{black}{(a) and (b) MP components $P_{M_x}$ and $ P_{M_y}$, (b) and (d) Spin polarization $P_{S_x}$ and $ P_{S_y}$ versus the lattice sites $n$ in the CHM. (e) $P_{M_y}$ and (f) $P_{S_z}$ versus $n$ in CHM. Note that in (a)-(d) and (f), the spin quantization axis is perpendicular to the helical axis while in (e), the spin quantization axis is parallel to the helical axis. The inset figures represents the MP in the right-handed CHM versus the energy $E$ and $n$.}}
\label{fig3}
\end{figure*}

The evolution of the lowest energy versus $B_z$ and the related probability distribution of the zero-energy states in the lattice sites of CHM are demonstrated in Fig. \ref{fig2}(c). We find that when $B_z$ reaches a critical value, the energy gap closes well and a pair of MZMs with zero energy appear to be localized at both ends of the CHM and are topologically protected against perturbations, verifying that the CHM can be worked as a new material platform to generate MZMs. To uncover the unique nature of this kind of MZMs, we turn to examine their lattice-dependent Majorana polarization (MP) \cite{Sticlet2012, Sedlmayr2015_1,Sedlmayr2015_2,Maska2017_2} and its components. Here the MP may be defined by any eigenstate $|\psi\rangle$, 
\begin{equation}
P_{M,n} (\omega) = \langle \psi | \mathcal{C} \hat{r} | \psi \rangle = \sum_{m=1}^{4\mathcal{N}}\sum_{\sigma} \delta(\omega-E_m) \sigma_{z} 2 u_{n \sigma}^{m} v_{n \sigma}^{m},
\label{eq4}
\end{equation}
where $\mathcal{C}$ and $\hat{r}$ are the particle-hole operator and the projection operator, $u_{\sigma}^{m} v_{\sigma}^{m}$ reflects particle-hole overlap, as they represent the electron and hole parts of the wave function respectively, and $E_m$ is the $m$th eigenvalue of $\mathcal{H}_{\text{BdG}}$. Naturally, we may express the components of the MP vector in Majorana space $(P_{M_x},P_{M_y}) = (\text{Re}[P_{M}(0)],\text{Im}[P_{M}(0)])$ \cite{Sticlet2012}. Firstly, the spatial distribution of $P_{M_x}$ as a function of $B_z$ is calculated and illustrated in Fig. \ref{fig2}(d). It clearly demonstrates that when $B_z$ reaches the critical magnetic field for the topological nontrivial phase, $P_{M_x}$ rapidly increases from 0 to a finite value, and these nontrivial states are nearly located at both ends of CHM, confirming further the existence of MZMs. More importantly, at the 1st and the $\mathcal{N}$the lattice sites, the MP display completely opposite values as drawn in the inset of Fig. \ref{fig2}(d), indicating that the MP indeed occurs in CHM and this new freedom of degree in MZMs can be served as a good indicator for identifying the related topological phase transitions. Moreover, as the poloidal angle difference $\Delta\theta$ is changed to $-\Delta\theta$, that is, the handedness of CHM is changed from the left to the right one, both components of MP $P_{M_x}$ and $P_{M_y}$ changes to opposite values accordingly as illustrated in Fig. \ref{fig3}(a) and \ref{fig3}(b) respectively, indicating that the MP generated here is tightly connected with the structural chality of CHM. Thus we refer to this kind of MP as the chirality-induced MP (CIMP) and the structural chirality can be worked as a new effective way to manipulate MZMs and the related topological quantum computing.

\begin{figure*}
\includegraphics[width=1.8\columnwidth]{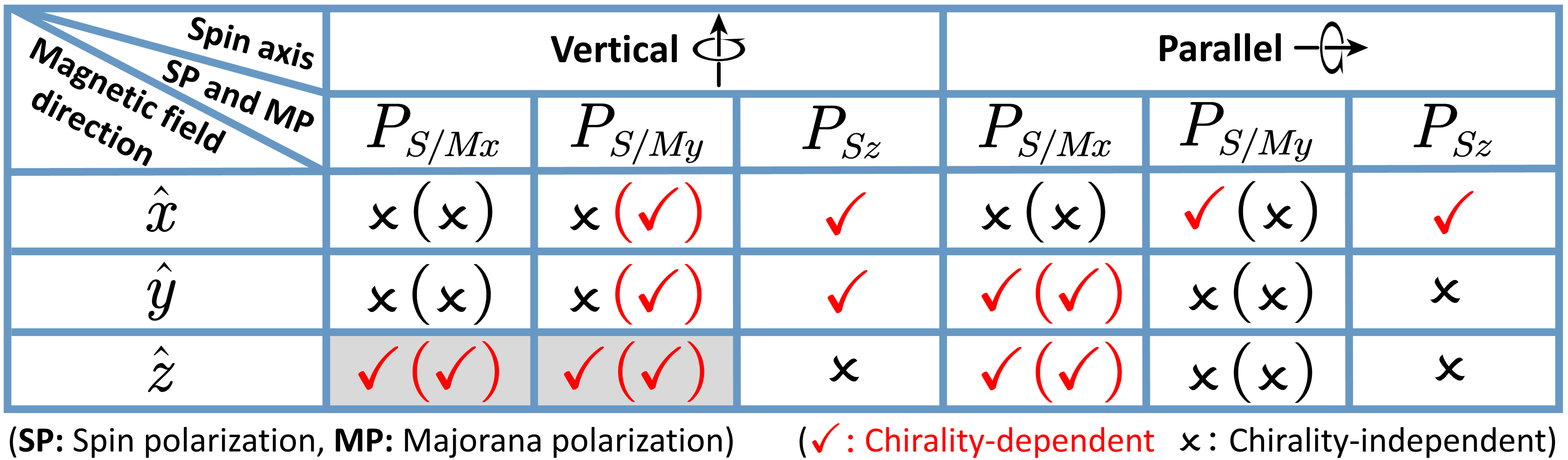}
\caption{\textcolor{black}{The relationship between the different components of spin polarization and MP  with chirality varies under different spin quantization axes and magnetic field directions.}}
\label{fig5}
\end{figure*}
To further understand the underlying physics in CIMP, we may explore its relation with the CISS in CHM. The components of the local chirality-induced spin polarization ($P_{S_i}$) versus the lattice site $n$ in CHM are given as 
\begin{equation}
P_{S_i, n}(\omega) = \sum_{m=1}^{4\mathcal{N}} \delta(\omega - E_m) \langle \Psi_n^{m} |  \frac{\tau_0 + \tau_z}{2} \otimes \sigma_i | \Psi_n^{m} \rangle,
\end{equation}
here $\tau$ are the Pauli matrices acting in the particle-hole spaces. We find that when the spin quantization axis and the direction of $B_z$ both perpendicular to the helical axis, as adopted in the calculations on CIMP (see the inset in Fig. \ref{fig3}(c)), both spin-polarized components $P_{S_x}$ and $P_{S_y}$ display as the same polarization phenomena as those occurring in $P_{M_x}$ and $P_{M_y}$ as illustrated in Figs. \ref{fig3}(b) and \ref{fig3}(d), indicating that the CIMP is tightly correlated to the transverse spin polarization. Thus the chirality-induced spin-polarized density of states meansurement can be worked as an effective way to detect the presence of MZMs in chiral materials. In particular, STM experiments can use selective equal-spin Andreev reflection spectroscopy to test the intrinsic polarization of Majorana quasiparticles \cite{He2014, Sun2016, Maska2017_1}. As a result, the spin-related component in CIMP is similarly represented by the spin-dependent zero-bias differential conductance.  \cite{Glodzik2020}
\begin{equation}
\lim_{V \to 0} \frac{dI_{i}^{\sigma}(V)}{dV} \simeq \frac{4e^2}{h} \left| 2u_{i\sigma} v_{i\sigma}  \right|^2.
\label{szbcp}
\end{equation}
Consequently, the experimental observation of the exchange of the two spin ZBCP values in different chiral molecules can be realized only when both components of CIMP reverse with the change in chirality.
% Therefore, only when both components of CIMP reverse with the change in chirality can the experimental observation of the exchange of the two spin ZBCP values in different chiral molecules be realized.

It is noted that in the conventional CISS experimental measurements, the electrons injected parallely to the helical axis of chiral molecules usually exhibit high (longitudinal) spin polarization. In our calculation on CIMP, however, as the spin quantization axis is changed to be parallel to the helical axis (see the inset of Fig. \ref{fig3}(e)), $P_{M_y}$ displays a Majorana state unrelated to the structural chirality as illustrated in Fig. \ref{fig3}(e), which is the main reason for us to perform our calculations using the spin quantization axis perpendicular to the helical axis, although $P_{S_z}$ is also free from chirality (see Fig. \ref{fig3}(f)). All scenarios of the relationship between spin polarization and MP with chirality under various spin quantization axes and magnetic field directions are shown in Fig. \ref{fig5}. On the other side, although the topological phase in our model exists in a wide range of parameter spaces, while applying an external magnetic field may cause other detrimental effects, such as to suppress the superconductivity by breaking Cooper pairs and to reduce the SC gap \cite{Pan2024, Liu2017}. In what follows, we may realize the CIMP in the superconducting interlinked-CHM chain in the absence of any magnetic field to verify the robustness and universality of the CIMP in chiral materials.

\begin{figure*}[t]
\includegraphics[width=1.9\columnwidth]{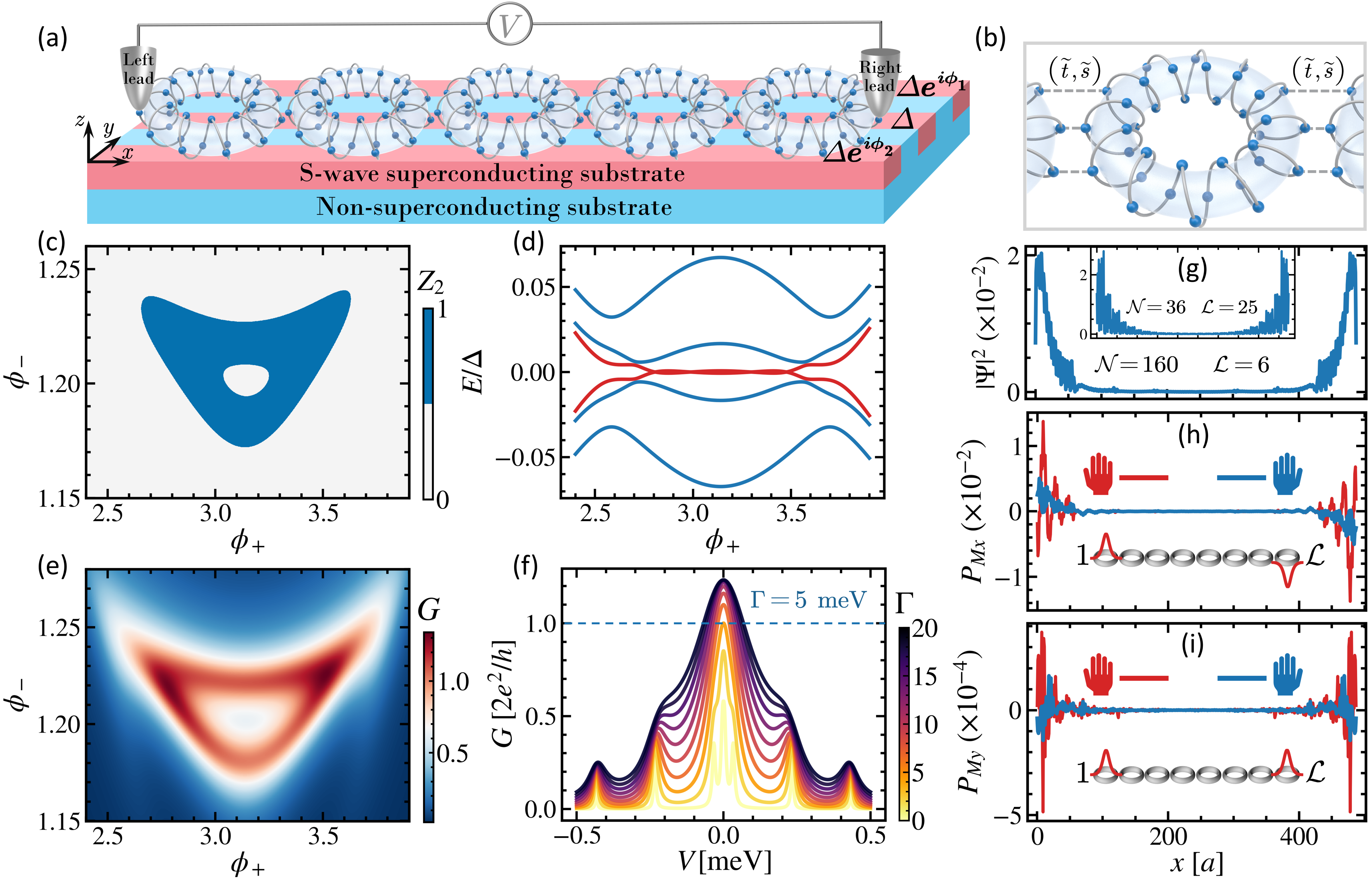}
\caption{\textcolor{black}{(a) A series of interlinked CHMs adsorbed onto a heterostructure composed of non-superconducting and phase-biased superconducting substrates and connected with two nonmagnetic leads. (b) An enlarged figure to show the intermolecular hopping integral ($\tilde{t}$) and the SOC strength ($\tilde{s}$).
(c) Topological Phase diagram by the $Z_2$ number versus $\phi_+$ and $\phi_-$. (d) Energy spectrum versus $\phi_+$ by fixing $\phi_- = 1.21$ and $\mu = 1.47$ meV. (e) ZBCP versus $\phi_+$ and $\phi_-$. (f) Differential conductance $G$ versus the applied bias $V$ for different $\Gamma$. Here the height of the ZBCP associated with the MSMs is only quantized to $2e^2/h$ at $\Gamma = 5$ meV. (g) Wave function distributions $\left| \Psi \right|^2$ of MZMs for $\mathcal{N} = 160$, $\mathcal{L} = 6$ and $\mathcal{N} = 36$ (inset), $\mathcal{L} = 25$. The wave function is integrated in the $y$-direction and displays along the $x$-direction. (h) $P_{M_x}$ and (i) $P_{M_y}$ versus the chain length $x$.}}
\label{fig4}
\end{figure*}

{\color{blue}\emph{Extending of CIMP to an interlinked CHM chain.}} To realize the above idea, a series of closed CHMs are interlinked each other using some experimental techniques such as hierarchical folding \cite{Qi2018} to form a quasi-1D helix-ring chain and meanwhile, these connecting CHMs are adsorbed onto a phase-biased non-SC/$s$-wave-SC heterojunction, and coupled by two nonmagnetic metal electrodes at two single lattices at both ends, as drawn in Figs. \ref{fig4}(a) and \ref{fig4}(b). To describe this interlinked-CHMs chain, a tight-binding Hamiltonian model is given as
\begin{equation}
\begin{aligned}
\mathcal{H}_{0} = \sum_{n=1}^{\mathcal{N}} \Bigg[ & \sum_{l=1}^{\mathcal{L}} \varepsilon_{nl} c_{nl}^\dagger c_{nl} 
+ \sum_{l=1}^{\mathcal{L}-1} \sum_{j=0}^{\mathcal{J}} \tilde{t} c_{n_j^\pm l}^\dagger c_{\tilde{n}^\mp_j,l+1} \\
& + \sum_{l=1}^{\mathcal{L}} t c_{nl}^\dagger c_{n+1,l} + \text{H.c.} \Bigg],
\label{eq7}
\end{aligned}
\end{equation}
where $c_{nl}^{\dagger}=(c_{nl \uparrow}^{\dagger}, c_{nl \downarrow}^{\dagger})$ is the creation operator for electrons at the $n$th lattice in the $l$th CHM, $\mathcal{L}$ is the total length of the CHMs chain, $\mathcal{J}$ is the maximum order of intermolecular coupling paths, and $c_{\mathcal{N}+1,l}^\dagger \equiv c_{1l}^\dagger$.  $\tilde{t}$ denote the intermolecular hopping integral. $n_j^\pm$=$\mathcal{N}/2 \pm\mathcal{M}$$\times{j}$+1 and $\tilde{n}^\mp_j$=$\mathcal{N}\delta_{1,\operatorname{sgn}(\pm j)}$$\mp \mathcal{M}\times{j}$+1 represent the lattice site in the $l$th and ($l$+1)th CHM respectively, as shown in Fig. \ref{fig4}(b). By a similar way, the chirality-induced SOC can be described by the following Hamiltonian
\begin{equation}
\begin{aligned}
\mathcal{H}_{\text{SO}} = -&\sum_{n=1}^{\mathcal{N}} \Bigg[\sum_{l=1}^{\mathcal{L}-1} \sum_{j=0}^{\mathcal{J}} i\tilde{s}\sigma_y c_{n_j^\pm l}^\dagger c_{\tilde{n}^\mp_j,l+1} \\
& + \sum_{l=1}^{\mathcal{L}} 2is\cos (\theta_{n}^{-}) \cos(\varphi_{n}^{-}) \sigma_{n}^p c_{nl}^\dagger c_{n+1,l} + \text{H.c.} \Bigg],
\label{eq8}
\end{aligned}
\end{equation}
where $\tilde{s}$ denotes the strength of the intermolecular SOC, and $\sigma_{n}^p$ is calculated by setting the spin quantization axis parallel to the helical axis as $\sigma_{n}^p=\cos \theta _{n}^{+}\sin \beta _{n} \sigma_x$+$(\sin \varphi_{n}^{+}\cos \beta _{n}$-$\sin \theta _{n}^{+}\cos \varphi _{n}^{+}\sin \beta _{n}) \sigma_y-(\sin \theta _{n}^{+}\sin \varphi _{n}^{+}\sin \beta _{n}$+$\cos \varphi _{n}^{+}\cos \beta _{n}) \sigma_z$. Considering further the s-wave SC substrate, we may replace $\mathcal{H}_{c}$ in Eq. (\ref{eq2}) by $\mathcal{H}_{0}+\mathcal{H}_{\text{SO}}$. Note that the SC pairings potentials $\Delta_{1}\sim\Delta_{3}$ of three particular regions $S_1\sim{S_3}$ drawn in Fig. \ref{fig4}(a) are assigned respectively as $\Delta e^{i\phi_1}$, $\Delta$ and $\Delta e^{i\phi_1}$, while zero in other regions. In every CHM, two key structural parameters are adopted as $r_0 = 1.9$$\textup{~\AA}$ and $r_1 =\mathcal{N}h/2\pi$, referring to the circular $3_{10}$ helical protein molecule with $\mathcal{N} = 160$, $h=2.0$$\textup{~\AA}$. Because the bare MZMs are delocalized in every single CHM, we may exhibit MZMs through trajectories interference \cite{Creutz2001}, that is, two neighbouring CHMs should be connected by multiple hopping paths. Therefore, we take $\mathcal{J} = 1$ model in which three paths are constructed for electron hopping between two adjacent CHMs. Moreover, the \emph{NN} hopping $\tilde{t} = 0.1 \, \text{eV}$, and the SOC is estimated to be $\tilde{s}=10\, \text{meV}$. The chemical potential in the CHMs is set to $\mu = 1.47 \, \text{meV}$ and the total chain length is set as $\mathcal{L} = 6$. The Nambu spinor is expressed as
$\Psi = \bigoplus_{n=1}^\mathcal{NL} \Psi_n$ and other parameters are adopted as above.

We plot the $\mathbb{Z}_2$ invariant of the CHMs proximity-coupled by a phase-biased superconductor versus the phase differences $\phi_+$$=(\phi_1+\phi_2)/2$, $\phi_-$$=(\phi_1-\phi_2)/2$ in Fig. \ref{fig4}(c) to show its topological phase transition. The plot shows that the nontrivial region ($\mathbb{Z}_2$=1) exhibits symmetry along the $\phi_+ = \pi$ axis. It is worth noting that the two key factors driving the system from a trivial to a nontrivial topological phase are the inherent closed trajectory structure with a non-zero AC phase ($\pi$ Berry phase) \cite{Aharonov1984,Aronov1993} arising from the chirality-induced SOC and the winding of the SC phase \cite{vanHeck2014,Fu2008,Lesser2021PRB}. This advantage, which allows for the induction of MZMs without the need for an external magnetic field, is not present in the previously studied linear open helical molecules. To confirm this nontrivial feature, we choose a typical path $l_1$ with $\phi_- = 1.21$ in Fig. \ref{fig4}(c) and along this path, the related lowest-lying energies versus $\phi_+$ are drawn in Fig. \ref{fig4}(d). It clearly demonstrates that with increasing $\phi_+$, an open energy gap at the zero energy is closed and then reopened again symmetrically, perfectly replicating the phase transition process along the path $l_1$. More than this, the ZBCP calculated in the phase region ($\phi_-$, $\phi_+$) displaying in Fig. \ref{fig4}(e) are also well consistent with the phase diagram given in Fig. \ref{fig4}(c), confirming the existence of topological SC in this interlinked-CHMs chain. It is evident that the ZBCP is enhanced with the increasing of $\Gamma$ from 0 to 20 meV, characterized by a quantized conductance of $2e^2/h$ observed for $\Gamma = 5$ meV. 

As expected, the real-space distributions of the wavefunctions along the $x$-direction at both ends of two representative CHM chains with $\mathcal{N} = 160$, $\mathcal{L} = 6$ and $\mathcal{N} = 36$, $\mathcal{L} = 25$ show clearly MZMs (see Fig. \ref{fig2}(g)) and when $\mathcal{N}$ is increased sufficiently, the MZMs are localized to be confined within two single CHMs in both ends and independent of the CHMs chain length, which is beneficial for experimental observation. More importantly, the both MZM components $P_{M_x}$ and $P_{M_y}$ display a chirality-dependent characteristic. That is to say the CIMP defined above also occurs in this interlinked-CHMs chain coupled with the $s$-wave SC heterostructure without any external magnetic field, which provides another advantage for this kind of MZMs to be test in experiments.

{\color{blue}\emph{Conclusion.}} To summarize, we have successfully proposed a new kind of material plateau, i.e., chiral materials, to generate MZMs. We found that the MZMs exist in both ends of opened CHM as it is absorbed on a $s$-wave SC in the presence of external magnetic field. Inspiring, the MZMs display opposite MP for the left-handed and right-handed configurations, indicating that we have realized the MZMs associated to the structural chirality of materials for the first time and thus, referred to this phenomenon as CIMP. Moreover, the CIMP is also associated to the chirality-induced spin polarization. Thus we have proposed that the structural chirality and the related CISS effect in chiral materials can be a new and effective way to detect and regulate the MZMs. Importantly, the CIMP also occurs in the interlinked-CHMs chain adsorbed onto a phase-biased non-SC/$s$-wave-SC heterojunction without any external magnetic field, verifying well the robustness and universality of CIMP.

{\color{blue}\emph{Acknowledgements.}} This work is supported by the National Natural Science Foundation of China with grant No. 11774104, 11504117, 11274128 and U20A2077, and partially by the National Key R\&D Program of China (2021YFC2202300).

\end{document}